%% file: main.tex
\documentclass[12pt]{article}
\usepackage{amssymb}
\usepackage{hyperref}
\usepackage{pgfplots} 
\usepackage{ulem} 
\usepackage{makecell}
\usepackage{comment}
\usepackage{caption}

\providecommand{\keywords}[1]
{
  \small	
  \textbf{\textit{Keywords---}} #1
}

\begin{document}

\title{Empirical Assessment of the Perception of Software Product Line Engineering by an SME \\before Migrating its Code Base
}
\author{Thomas Georges$^{1,2}$, Marianne Huchard $^{2}$, Mélanie König$^{1}$, 
\\Clémentine Nebut $^{2}$, Chouki Tibermacine $^{3}$\\
\small $^{1}$ LIRMM, Univ Montpellier, CNRS, Montpellier, France\\
\small $^{2}$ ITK -Predict \& Decide, Montpellier, France\\
\small $^{3}$ IRISA, University of Southern Brittany, Vannes, France
}

\date{}

\maketitle   

\begin{abstract}
Migrating a set of software variants into a software product line (SPL) is an expensive and potentially challenging endeavor. Indeed, SPL engineering can significantly impact a company’s development process and often requires changes to established developer practices. The work presented in this paper stems from a collaboration with a Small and Medium-sized Enterprise (SME) that decided to migrate its existing code base into an SPL.
In this study, we conducted an in-depth evaluation of the company’s current development processes and practices, as well as the anticipated benefits and risks associated with the migration. Key stakeholders involved in software development participated in this evaluation to provide insight into their perceptions of the migration and their potential resistance to change.
This paper describes the design of the interviews conducted with these stakeholders and presents an analysis of the results. Among the qualitative findings, we observed that all participants, regardless of their role in the development process, identified benefits of the migration relevant to their own activities. Furthermore, our results suggest that an effective risk mitigation strategy involves keeping stakeholders informed and engaged throughout the process, preserving as many good practices as possible, and actively involving them in the migration to ensure a smooth transition and minimize potential challenges.
\end{abstract}


\input{1.introduction}

\input{2.Contexte}

\input{3.Methodology}

\input{4.Results}

\input{5.Discussion}

\input{6.ValidityStudy}

\input{7.RelatedWork}

\input{8.Conclusion}

\bibliographystyle{alpha}

\newcommand{\etalchar}[1]{$^{#1}$}

\end{document}

%% file: 1.introduction.tex
\section{Introduction}

A Software Product Line (SPL) can be derived from an existing code base containing a set of related software systems that share common characteristics, with the goal of enabling the rapid and flexible creation of new software products \cite{DBLP:conf/icsm/FischerLLE14,DBLP:conf/splc/MartinezZBKT15,DBLP:conf/splc/MartinezAZ17}.
This extractive approach, which consists in migrating an existing code base into an SPL, is inherently complex and time-consuming.
Moreover, it may have a significant impact on well-established development processes and practices within the organization managing the code base.

This paper presents a preliminary step toward the concrete implementation of such an extractive process.
The work is part of a joint project with ITK - Predict and Decide, a small and medium-sized enterprise (SME) that decided to migrate its code base into an SPL.
Our role as a research team was to support and guide this migration process.

The ITK code base is organized around a platform called \texttt{Cross}, which is an extended framework in which software reuse is already a common practice.
This framework resembles an ad hoc SPL, yet it lacks the models and processes that characterize a traditional SPL.

To assist the company’s developers, we planned a gradual transition toward a progressive reorganization of both their code base and their engineering practices.
Our first task was to analyze how this transition was perceived by the staff and to provide recommendations to facilitate the migration process.
The objective of this paper is to report the conclusions of this preliminary study.

We conducted an evaluation of the company’s current practices and of its employees’ perceptions regarding SPL engineering.
This evaluation was implemented through a series of structured interviews with a representative group of participants.
The interview design follows the methodological recommendations of Kitchenham et al.~\cite{DBLP:books/sp/08/KitchenhamP08} and is presented in detail in this paper, together with the obtained results.
The interviews were tailored to the various categories of participants and conducted with a focused group representing different roles within the organization.
Specifically, about 15\% of the SME’s staff participated in the study, including 50\% IT specialists, 30\% product owners (POs), and 10\% agronomists, along with all UI/UX designers.
This composition ensured coverage of nearly all stakeholders involved in the migration project.
Other employees were not included, as they were engaged in unrelated projects (such as breeding or carbon sequestration) or administrative roles (HR, finance), and therefore not directly concerned by the migration.
The interviews explored multiple perspectives: perceptions of current best practices, expected benefits of the migration, and perceived risks and concerns associated with adopting an SPL.

The results show that most participants were positive about the SPL migration and expressed a strong willingness to be actively involved in the process.
Developers were the most enthusiastic group, perceiving the migration as both a technical and managerial challenge.
Actors closer to business activities expressed more reservations, emphasizing potential risks such as increased maintenance complexity or threats to creativity, particularly in UI/UX-related work.

Beyond the qualitative insights, the interpretation of these results led to the formulation of practical guidelines to support ITK’s SPL migration.
These findings also provide a foundation for defining an agile SPLE (Software Product Line Engineering) process aimed at mitigating the identified risks and concerns.

Following the framework proposed by Salo and Abrahamsson~\cite{DBLP:conf/profes/SaloA04}, our empirical study is structured into three main components:
\begin{itemize}
    \item Goal Setting (Section~\ref{sec:context}):
This stage defined the purpose of our investigation, assessing perceptions of SPL migration.
Four key topics guided the interview design: current best practices, perceived benefits, perceived risks, and concerns regarding the migration.

\item Data Collection (Sections~\ref{sec:methodo} and~\ref{sec:results}):
To ensure reliability, data collection was carefully planned and implemented through interviews aligned with the four identified topics.

\item Data Interpretation (Sections~\ref{sec:discussion} and~\ref{section:validity}):
Interpretation was conducted with particular attention to avoiding bias in favor of SPL adoption.
Long and open discussions were held with interviewees, and their anony\-mized responses are made publicly available~\cite{georges_2025_17598548}.
\end{itemize}

The remainder of this paper is organized as follows.
Section~\ref{sec:context} presents the research context and questions.
Section~\ref{sec:methodo} describes the methodology.
Section~\ref{sec:results} discusses the results.
Based on these results, Section~\ref{sec:discussion} outlines recommendations for preparing the SPL migration.
Threats to validity are discussed in Section~\ref{section:validity}.
Related work is presented in Section~\ref{section:relatedwork}, and
Section~\ref{sec:conclusion} concludes the paper with a summary of our findings and directions for future work.

%% file: 2.Contexte.tex
\section{Context and Research Questions}
\label{sec:context}
    
    This section presents the context that guided the development of research questions for the interview design in this study. It begins with the background (Sect. \ref{subsec:background}), followed by the context (Sect. \ref{subsec:context}), and concludes with the research questions formulated for the interview design (Sect. \ref{subsec:rq}).

    \subsection{Background}
        \label{subsec:background}

        Software Product Lines are now a well-established way to efficiently produce highly-configurable software products \cite{DBLP:books/daglib/0015277}. A software product line is a software supply chain roughly composed of several components. Variability models expose the available product features and the authorized ways of combining them in product configurations. An asset base provides the concrete software pieces that implement the features. Tools that are part of the SPL enable the selection of features to make a concrete configuration, and then to generate the concrete product from the configuration by assembling assets from the base. Such an approach becomes worthwhile when a sufficient number of configurations need to be developed. This is not obvious to plan for a company which begins with a new product kind. Thus many companies often develop first a family of similar software systems before deciding to invest effort in migrating their software development infrastructure to a product line.
        Despite successful past experiences reported in the literature~\cite{DBLP:conf/splc/AbbasJLESS20}, migrating an existing software family into an integrated SPL platform is still challenging and organizations may be reluctant to adopt the approach. Hesitations often arise from the lack of standard procedures guiding the process, as well as doubts on the cost/benefit ratio. Thus, although companies own large bases of well-documented code, they still manually build tailored applications from the base code to meet their clients' requirements, in particular with a clone-and-own strategy.

    \subsection{Context of this work}
        \label{subsec:context}
        Our partner, ITK -- \textit{Predict and Decide}, is an SME with 120 collaborators at the time of writing this paper, among which 19 agronomists and 29 developers including 2 UI/UX designers and 4 system administrators. One-third of the company employees are Tech-oriented collaborators. The company develops its systems in an agile way with a well-functioning development and delivery process. The applications that are developed assist farmers in decision-making by enabling them to predict diseases and crop win for their plants and livestock. The core of these applications is a set of simulators built by the agronomists. Then the IT teams develop additional code which collects data from IoT devices or online services in order to feed the simulators and display their output data in dashboards. The back-end of the software is constructed using Kotlin, while the front-end is developed with Angular, with Typescript being utilized prominently. The server operates on Spring Boot, and the database is hosted by PostgreSQL. Agronomists use Python or MATLAB to build the simulators.

        The IT team uses a central platform, named Cross, which is a large framework, to build applications code. Initially, the platform was designed to generate \textit{Proof of Concepts}. Over time, this purpose has evolved, and the platform has been transformed into a foundation for end products. There are currently 7 variants in the family for 3 Product Owners. New variants are tailored to specific client needs, with a new variant being introduced approximately once per year.

        The levels of the \textit{Architecture Dimension}, as outlined in the \textit{FEF framework}~\cite{DBLP:books/daglib/0018329}, deal with the technical means to build the software and determine the technical realization of the products within the software product line. At Level 1, the architectures are limited to single systems, lacking any visible focus on reuse. Moving to Level 2, there is a shift towards reusing third-party infrastructure, with defined common software components like middleware, but no formal incorporation of domain-specific assets. Progressing to Level 3, domain commonality is captured and implemented into a software platform, offering a reference architecture for all applications, primarily centered around the platform utilization. However, this configurable platform lacks support for handling configuration variability. Advancing to Level 4, both domain commonality and variability are systematically captured, and a reference architecture is specified to cover the entire software product line. Domain assets are designed to support the derivation of products, with explicit attention given to variability management within the software product line architecture. Finally, at Level 5, the reference architecture takes a dominant role, with application architectures only marginally diverging from it. Automated derivation of products becomes feasible through the use of scripts, tools, and high-level languages, while application development primarily revolves around configuring within the reference architecture's boundaries, leading to the possibility of automated product configuration.

        This platform is 5-6 years old. According to the criteria of \textit{FEF framework}~\cite{DBLP:books/daglib/0018329}, the ITK's home-made platform is a \textit{``Level~3: Software Platform``} with a reference architecture common to all products. The current architecture contains a collection of common assets and explicit variation points for new variants are defined. The migration goal is to reach an architecture level where domain commonality and variability are captured, and a reference architecture is specified for the complete software product line. This includes systematic and managed reuse based on an asset repository, explicit variability in the assets, and an explicit reference architecture that governs the permissible variations in application architectures, while incorporating numerous quality solutions in the software product line architecture. Variability management is addressed explicitly in the software product line architecture, determining allowed configurations for application architectures and defining variation points and constraints for most of these variations, including rules that application-specific variants must adhere to. The ultimate objective is providing a software configuration for an automated variants synthesis.
        
        According to the FEF framework~\cite{DBLP:books/daglib/0018329}, at the \textit{organizational} level, ITK operates at the fourth level. This implies the synchronization and coordination of domain engineering and application engineering. The project development is carried out through a strong collaboration between agronomists and software developers.
        The migration decision and the collaboration with our research team was taken in a collegiate discussion with the project leaders and stake-holders.
        Before starting the design of interviews, we did not yet establish a well-defined research methodology for the rest of the project. We wanted first to assess the perceptions of the company's actors regarding the migration to SPL. Depending on the results of the interviews, we will engage in the elaboration of a concrete methodology for the migration project.

        The mentioned arguments are to maximize code reuse and further accelerate the delivery of new applications. The goal is to migrate the code base into an SPL, at a higher architecture level (Level 4 or 5 in~\cite{DBLP:books/daglib/0018329}). In order to smoothly make this transition, we thus decided to first conduct interviews with the collaborators of the company to evaluate their perception of SPL engineering. The interviews have been designed in such a way to answer research questions presented in the following section. Another round of interviews is planned at the end of the project to refine the guidelines for a successful migration.

        The following migration phases will involve variability analysis, starting by establishing a process for feature localization to synthesize feature models. Then, the study of interactions and dependencies among the features will lead to achieving a comprehensive system description.

    \subsection{Interview Design Research Questions}
        \label{subsec:rq}
        We built and conducted the interviews in order to identify what is the perception of current practices, benefits, risks and fears before the migration to a SPL.

        \paragraph{RQ1} \textbf{What is company's actors perception of current development good practices?}
            Through this question we want to highlight perception of the current good development practices of the company's actors.
            
        \paragraph{RQ2} \textbf{What is company's actors perception of benefits of the future SPL?}
            The idea here is to discuss with the company's actors the added value of the future SPL from an application development perspective.
            
        \paragraph{RQ3} \textbf{What is company's actors perception of risks of the future SPL?}
            Once discussing the benefits, we want here to analyze the risks of the future SPL in the daily work of the company's actors.

        \paragraph{RQ4} \textbf{What is company's actors fears induced by the previously identified risks?}
            This question enables to assess the fears felt by the company's actors about the consequences of the migration project.  
            Fears refer to the apprehensions or concerns that company actors may have regarding the transition process. Such fears might arise from the uncertainty of the outcome, potential challenges during the migration process, and the need for proper planning and mitigation strategies to address the identified risks effectively.

        We elaborated a methodology that guided the design and subsequent conduction of the interviews, in order to answer these questions.

%% file: 3.Methodology.tex
\section{Methodology}
\label{sec:methodo}
    Interviews are one of the direct inquisitive data collection techniques to conduct field studies and gather opinions \cite{DBLP:books/sp/08/SingerSL08}. 
    We chose them rather than a written questionnaire or a focus group. 
    Compared to a focus group, the one-to-one aspect of an interview avoids respondents influence each other. 
    Compared to a written questionnaire, interviews require more respondent commitment and time as they necessitate meetings with the interviewer. In return, they are interactive and flexible, as the interviewer may clarify the questions in case of ambiguity, and extend the discussion beyond the initial question set. 
    We explain in this section how we have chosen the participants, and then how we designed and conducted the interviews.
    
    Our empirical assessment follows the principles presented in ``Empirical Research Methods in Software Engineering''  \cite{DBLP:conf/esernet/WohlinHH03}. The \textit{goal setting} component consists of formulating well-defined research objectives, which is required to conduct effective empirical studies. Defining the study's purpose as the assessment of perception surrounding SPL migration, as we did, mirrors the book's emphasis on establishing clear research goals. The \textit{data collection} phase, involving the careful design of interviews centered around specific topics, has to be put into perspective with the book's discussions on selecting appropriate data collection methods tailored to the research context. In particular, we did our best to apply the book recommendations on rigorous data collection planning for valid and reliable results. The \textit{data interpretation} refers to the significance of unbiased analysis. Engaging in thorough discussions, attentive listening, and accurate transcription to extract insightful qualitative data aligns with the book's guidance on minimizing interpretation biases. 

    The methodology section is composed of the participants presentation (Sect. \ref{subsec:participants}), the interview design (Sect. \ref{subsec:interviewdesign}), our approach to conduct interviews (Sect. \ref{subsec:conductinterviews}), and the interview analysis method (Sect. \ref{subsec:analysis}).

    \subsection{Participants}
        \label{subsec:participants}
        We selected the participants of the interviews to have a representative population \cite{DBLP:books/sp/08/KitchenhamP08}.
        They cover the different types of actors of the project, different levels of experience in the project, and different educational backgrounds.
        Four types of actors participate in the project: Product Owners (PO), Developers (IT), Agronomists (AGRO) and User Interface/Experience (UX) designers. The actors have different educational backgrounds. Here we distinguish between persons holding an engineering degree from those holding a PhD.
        We also selected participants with different levels of experience in the project.
        Finally, we selected 16 participants to be representative, among them nine are IT who are working currently or who worked in the past on the project, two are Agronomists, three are POs and two are UXs, all of them working on the project.
        Table \ref{table:gradeJobAgeOfServ} shows educational background distribution of the participants per role and shows the levels of experience in the company.
 
            \input{gradeJobAgeOfServ}

        Motivation is important to have relevant answers \cite{DBLP:books/sp/08/KitchenhamP08}.  
        In this study, respondents knew that they will have a feedback and that their answers would be used to define a good strategy while implementing the SPL process.

    \subsection{Interview design}
        \label{subsec:interviewdesign}

        We chose the form of structured interviews with a list of questions. We did our best to follow the main recommendations of \cite{DBLP:books/sp/08/KitchenhamP08} with terms that can be understandable and unambiguous. Each question relates to one concept. We avoided sensitive questions or interrogations on old events. We also avoided questions that are inappropriate because the respondent has no sufficient knowledge to answer. This last aspect led us to consider part of the actor-driven interviews. 
        The interview questionnaire consists of three main sections. The first section is an introduction where the interviewee is requested to present her/him-self. The main section follows, which is further divided into two subsections - one focusing on the current process and the other on migration. Finally, there is a conclusion section that covers the discussion about the interview as a whole.
        
    \subsubsection{Actor-driven interviews}
        The actors involved in the project have varying perspectives and levels of proficiency in software engineering. To take this into consideration we used different variants of the questionnaire. 
        We knew that the ITs and UXs have the most technical profiles, the POs are the nearest to business and the agronomists have a technical profile (in agronomy and in programming the simulators), but quite far from code reuse concerns. The agronomists were involved in order to assess a different point of view on the migration project. 
        As specified in Table~\ref{table:questions}, the interview questionnaire comprises both common and specific questions. This implies that certain parts of the interview are exclusively intended for specific actors.
        
        \input{questions}
        
        The common questions are basic questions about the interviewee and their history in the company. The specific questions have variants corresponding to each role:
        \begin{itemize}
            \item For ITs: questions focus on the source code internal quality, such as reusability or redundancy.
            \item For POs: questions focus on the project management and team practices (e.g. agility practices).
            \item For UXs: questions focus on the front-end and interfaces development.
            \item For AGROs: questions focus on prediction simulators development and their collaboration with the ITs.
        \end{itemize}

    \subsubsection{Topic-driven interviews}
        We built our interviews on four main axes, to identify: (1) the current good practices, and, related to the migration (2) the expected benefits, (3) the risks and (4) the fears.

        We consider the \textit{current good practices} by asking the following questions:
        \begin{itemize}
            \item What do you think about the existing practices?
            \item What would be the points to keep?
            \item In terms of development by and for reuse, what is your general opinion on the code base and on the common base code itself?
        \end{itemize}
        In the last question, the statement ``In terms of development by and for reuse'' refers to the way software development has been conducted in the platform, whether developers emphasize on creating code with the intention of making it easily reusable in various contexts (development for reuse) and whether they maximize code construction by reusing and assembling existing code (development by reuse).

        The \textit{benefits} expressed about the new approach are deduced from the following questions:
        \begin{itemize}
            \item What do you think about a platform to develop (derive) similar software quickly?
            \item Do you see the interest of such an approach in the current processes of the company?
            \item Which good practices can be derived from this approach?
        \end{itemize}
        
        Asking these questions related to the benefits will potentially highlight the expected changes that will eventually eliminate the bad practices in the current development process (before migration).
        
        The possible \textit{risks} related to the considered approach are deduced from the following question:
        \begin{itemize}
            \item Which risks can arise from this approach on your current ways of working?
        \end{itemize}
        The \textit{fears} which limit the enthusiasm are deduced from question:
        \begin{itemize}
            \item Do you have any fears related to this kind of change?
        \end{itemize}
    
    \subsection{Conducting the interviews}
        
        \label{subsec:conductinterviews}
        We planned the way interviews were to be conducted as follows. We wanted interviews to be fluid, more like a discussion than just Q\&A sessions.
        
        The first part of the interviews, before asking the first question is a brief reminder of the principles of SPLE and the purpose of the questionnaire. Then the first questions are about the interviewed actor in the company. The purpose here is to understand the profile of the actor. We continue by general questions on the company development process. The next part of the interview contains more precise questions on the current development process of new products. This part aims at understanding what are the perceptions of the current strengths, the good practices and the state of the process.
        Discussing the current state of practice is the perfect transition to talk about a new approach which may be able to overcome the current weaknesses. The last part is dedicated to discuss the migration. Participants are encouraged to express their fears, the detected risks and the possible benefits. The two last questions aim at improving the interviews and concluding.
        
        All the interviews were conducted by the same interviewer who was the first author of this work. Each interview took between 30 minutes and 1 hour. They were appreciated and interviewees reported that they felt included in the migration project by the way. Each interview ended with the question: ``Was there a question you were expecting that I didn't ask?''. This question helps the interviewed person to think about all the questions, to complete what they wanted to share, and participate to improve the fluidity of the next interviews. The questions remain unchanged to ensure homogeneity in data analysis. The intent is to enable participants to think about all aspects without altering the interview questions while still actively participating in improving the fluidity of the interviews.
        The transcription was made by note-taking. Still following recommendations by \cite{DBLP:books/sp/08/KitchenhamP08}, the interviewer made his best to ensure that questions and exchanges do not have bias, such as influencing the respondent by the question wording, or the question order. He also took care about asking enough questions and tried not to influence the respondent.
        As a pilot, the first interview was made with our main collaborator in the company who supervises this research work with us, in order to improve the questionnaire and ensure the fluidity of the process. The data from this first interview were not considered in the interview analysis.

    \subsection{Interview analysis}   
    
        \label{subsec:analysis}
        We first qualitatively analyzed the results by carefully reading several times the interview summaries. During the interviews, we recorded the discussions, took handwritten notes, and later summarized the recorded content. This combined approach aims to ensure that all relevant information is captured and to minimize the risk of missing critical details, aligning with the common and recommended practice of recording and transcribing interviews in qualitative research. The analysis process was made by the research team, which included the person who conducted the interviews.

        Then we designed a set of metrics to conduct a quantitative analysis. These metrics are a combination of \textit{tallying responses} and \textit{frequency}:
        First, we counted \textit{how many current good practices are identified}, next \textit{how many benefits are identified?}, then \textit{how many risks are identified?} and finally \textit{how many fears are identified?}.
        To answer these questions we looked for similar formulations of answers and we interpreted interview's answers to extract simple words summarizing the ideas. For example, \textit{``The genericity pushes to factorize more and more''} is summarized with \textit{``genericity'' and ``factorisation''}.
        For each topic we first identified the frequency of generic expressed terms. Then we separated them per role -- the actor who expressed them.
        
        The words used to summarize the answers have been manually established, according to our knowledge about Agile methods and SPL Engineering. 
        These words have been \textit{a posteriori} validated, by identifying them in a set of large documents which are: The Agile Manifesto~\cite{beck2001agile}, Software engineering - principles and practice~\cite{DBLP:books/daglib/0020600} and SPL Engineering - Foundations, Principles, and Techniques~\cite{DBLP:books/daglib/0015277}. The words related to feelings and risks have rather been identified simply from a dictionary.
        
        The potential words were identified with a method composed of several steps. We first sorted words by frequency after having removed stop-words and punctuation. We selected the most meaningful words, then we browsed the documents to validate the context of use for each word. Our identified words were enriched to form compound words by adding adverbs, like \textit{``too complex''} instead of \textit{``complex''} to strengthen the meaning of the word.
        We compared the words deduced from the books with the words from the answers. First we checked if there was a match between them. If there was a negative match we looked for more suitable words. For instance, we replaced ``unusable'' with ``useless'' to better align with the identified words from books. If there was a positive match we concluded that the word is correctly aligned with the potential identified words.
        This process allowed us to refine and improve our synthesized answers with knowledge coming from books and get a better match with the language used in those books. Furthermore, this approach helped us to extract the most relevant and meaningful terms to improve our data analysis.

        To perform the frequency analysis and the similarity processing we developed a script in Python in ``Jupyter Notebook''. Once the terms from the book sorted by frequency we manually choose the meaningful words.
        
        Our data and results are available in an online repository:\\\href{https://gite.lirmm.fr/tgeorges/interviewartefact}{https://gite.lirmm.fr/tgeorges/interviewartefact}. 
       The repository includes the questions and corresponding answers, the analysis of the interviews, the topic lists, the notebook and several bar charts depicting responses.

%% file: gradeJobAgeOfServ.tex
\begin{table}[htb]
\centering
\caption{Roles (column 1), educational background (columns 2 and 3) and average number of years of service (column 4) of the interviewees.}
\begin{tabular}{|l|l|l|l|}
\hline
Role  & PhD & Engineer & \makecell[l]{Average \# years of service\\(standard deviation)} \\ \hline
IT    & 2   & 7        & 6.1~(1.5)                    \\ \hline
PO    & 2   & 1        & 5.7~(2.6)                      \\ \hline
UX    & 0   & 2        & 8.5~(2.1)                      \\ \hline
AGRO  & 2   & 0        & 6~~~(1.4)                      \\ \hline
TOTAL & 6   & 10       & 6.6~(2.2)                      \\ \hline
\end{tabular}
\label{table:gradeJobAgeOfServ}
\end{table}

%% file: questions.tex
\begin{table}[htb]
\centering
\caption{Number of questions common to all roles and additional questions for the specific roles.}
\begin{tabular}{|l|r|}
\hline
Common & 12 \\ \hline
IT     & 12 \\ \hline
PO     & 8  \\ \hline
UX     & 12 \\ \hline
AGRO   & 12 \\ \hline
TOTAL  & 56 \\ \hline
\end{tabular}
\label{table:questions}
\end{table}

%% file: 4.Results.tex
\section{Interview result}
\label{sec:results}

    In this section, we analyze the results per research question: current \textit{good practices} (Sect. \ref{subsec:cpg}), \textit{benefits} (Sect. \ref{subsec:benefits}), \textit{risks} (Sect. \ref{subsec:risks}) and \textit{fears} (Sect. \ref{subsec:fears}) perceived for the future SPL.
    Table~\ref{table:tableOverview} details the number of identified good practices, benefits, risks and fears that are further analyzed. The results are presented according to topics and roles, following the methodology that guides the design process.
    
    The responses given by the interviewees are evenly spread. When we analyze the responses based on the number of years of experience, both the least and the most experienced individuals did not yield a significantly different quantity of answers. For instance, when we asked ``Which good practices can be derived from this approach?'', developers with either 4 or 7 years of experience provided the same responses. Multiple bar charts depicting responses based on years of experience can be found within the repository accessible through the article.

        \input{tableOverviewResult}
        
    \subsection{Current good practices (RQ1)} 
        \label{subsec:cpg}
        This part of the interviews aimed at detecting the current good practices that are crucial to make them persistent in the new approach. 
        The most frequently cited practice is \textbf{Factorization and Reuse}. 
        This is supported by team interactions  
        where each new feature is discussed to determine if it is specific to one application or more generic and available for all applications. 
        \textbf{Innovation} is a predominant idea which motivates the team to keep their knowledge up-to-date. This is important to the participants and they have the possibility to regularly upgrade the code base with new features and framework updates. Innovation concerns technical IT aspects and advances in agronomy as well.
        \textbf{Fast} development is the basic idea of the project. It involves the capacity to produce a new application faster than the previous one.
        \textbf{Software Engineering spirit} combines all the principles detailed above, i.e. factorization, reuse, innovation and rapidity. The project was initially thought and designed like a common reusable code base. The proximity to the SPLE objectives helps to promote the migration project.
        We identified \textbf{7} different current good practices. We noticed that \textit{Factorization} and \textit{Reuse} are the two most frequently mentioned ideas with 9 and 8 occurrences, respectively (Table~\ref{table:tableCGPA}).
        Figure~\ref{fig:radarChartCGP} graphically highlights the distribution of the mentions.\\ 
        \begin{minipage}{0.47\linewidth}
            \includegraphics[width=1.\linewidth]{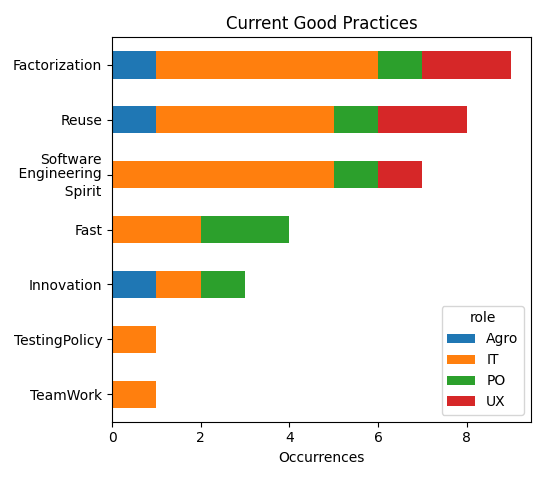}
            \captionof{figure}{The most frequently mentioned ideas about current good practices.}
            \label{fig:radarChartCGP}
        \end{minipage}\hfill%
        \begin{minipage}{0.5\linewidth}
        \input{tableCGPA}
        \end{minipage}
            
    \subsection{Benefits (RQ2)}
        \label{subsec:benefits}
        The perceived benefits are very important to know because their expression will be motivating for the whole team.
        \textbf{Customization} is one of the main characteristics of an SPL and was the most frequently mentioned idea by \textit{ITs}, \textit{UXs} and \textit{AGROs}. An SPL is perceived as a platform that is able to generate a mass customized software from a collection of existing software.
        The \textbf{rapidity}, mentioned by \textit{ITs}, \textit{POs} and \textit{UXs}, is the second most frequently mentioned idea. It corresponds to the capacity to quickly derive a new software.
        \textbf{Automation} has also been mentioned. It is the process of deriving new software automatically. \textit{ITs} and \textit{POs} linked automation and \textit{rapidity}.
        \textbf{Industrialization} was mentioned by \textit{ITs} and \textit{UXs}, meaning here the on-demand and automated production of applications at low cost.     
        Improving the \textbf{code quality} was mentioned only by \textit{ITs}. Code quality here takes into consideration code homogeneity, testability and reliability.
        \textbf{Selling (or rapid time-to-market) advantages} were mentioned by \textit{ITs} and \textit{POs} and grouped all characteristics allowing to improve the business part. It can refer, for example, to selling more software or answering quickly to a call for tender.
        These \textbf{7} benefits are highlighted in Figure~\ref{fig:radarChartbenefits}. We noticed that \textit{Customization}, mentioned 10 times, is the most frequently expressed idea, followed by \textit{Fast} (see Table~\ref{table:tableBenefitsA}).\\
        \begin{minipage}{0.43\linewidth}
                \centering
                \includegraphics[width=1.\linewidth]{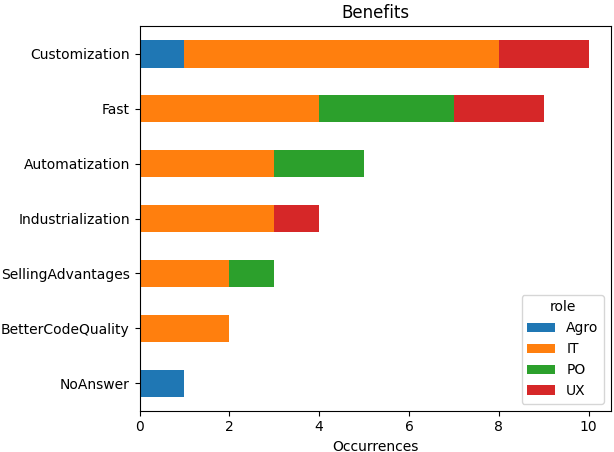}
                \captionof{figure}{The most frequently mentioned ideas about future benefits.}
                \label{fig:radarChartbenefits}
        \end{minipage}\hfill%
        \begin{minipage}{0.52\linewidth}
            \input{tableBenefitsA}
        \end{minipage}
        
    \subsection{Risks (RQ3)}
        \label{subsec:risks}
        Risks are possible weaknesses brought by the migration and are important to identify to be able to avoid or at least limit them.
        The most frequently mentioned risk is \textbf{Complex}, mentioned by the \textit{ITs} and the \textit{POs}. This perception comes from a misconception of Software Product Line Engineering (SPLE). For those new to the field, understanding the operation of an SPL and its associated methods can be challenging. 
        The second most frequently expressed idea is \textbf{the difficulty to maintain and evolve} the SPL by seeing it as a black box, which limits the possibility to make evaluations. That was mentioned by \textit{ITs} and \textit{UXs}. 
        Then, the \textbf{loss of features}, expressed by \textit{ITs} and \textit{AGROs}, is the inability to retrieve all the current features and even re-create the existing current applications.
        The \textbf{loss of creativity} was expressed by \textit{UXs}. It results from the lack of faith into the capacity of the SPL to evolve and the fact it could restrict the types of evolution.
        Expressed only by \textit{POs}, the \textbf{excess of side effects} is due to the factorization of features. A feature modification may impact all the applications using it, and could spread bugs.
        Finally, the \textbf{time-consuming} aspect of migration to an SPL was mentioned by \textit{POs} who were worried about the balance between the benefits and the amount of time needed to implement the SPL. 
        Figure~\ref{fig:radarChartrisks} shows the \textbf{7} identified risks. We noticed that \textbf{Complex} was the most frequently mentioned risk, followed by \textbf{Hard to maintain} (Table~\ref{table:tableRisksA}).\\
        \begin{minipage}{0.40\linewidth}
            \centering
            \includegraphics[width=1.\linewidth]{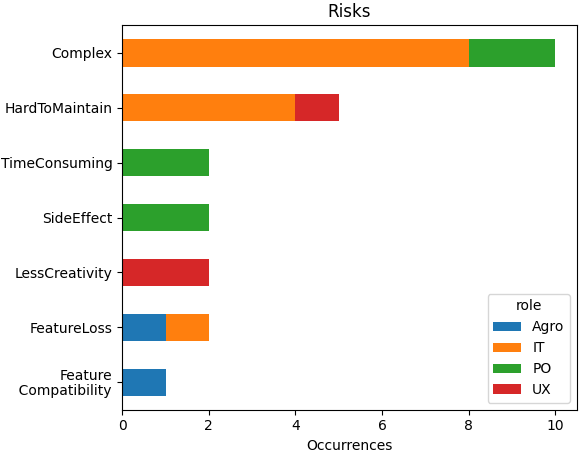}
            \captionof{figure}{The most frequently mentioned ideas about risks.}
            \label{fig:radarChartrisks}
        \end{minipage}\hfill%
        \begin{minipage}{0.55\linewidth}
        \input{tableRisksA}
        \end{minipage}
        
    \subsection{Fears (RQ4)}
    
        \label{subsec:fears}
        Fears are the concerns, the \textbf{subjective} negative feelings about migration, that are important to consider to reassure the actors. They are the perceived undesired possible consequences. 
        When answering the first two questions related to the benefits, our interviewees highlighted fears induced by the risks they have identified. Two key questions that have indirectly supported fear identification are:
        \begin{itemize}
            \item Do you see the interest of such an approach in the current processes of the company?
            \item What do you think about a platform to develop (derive) similar software quickly?
        \end{itemize}
        
        None of the \textit{UXs} expressed fears, contrarily to \textit{POs} who all expressed worries.
        The most frequently expressed fear is the creation of a  \textbf{useless} system (the SPL itself), too far from the reality and which prevents any evolution. 
        Another fear is the possible \textbf{excess of added complexity} of the project limiting the interest and involvement of the actors. 
        They also report \textbf{resistance to change}, a fear common to all new projects. 
        Actors are also concerned by the fact that the project may be more \textbf{expensive} than profitable.
        Finally, they worry that the result will be in  \textit{regression} compared with the current process.
        
       The \textbf{7} identified different fears are shown in Figure~\ref{fig:radarChartfears}. We noticed that the main fears of ITs are \textbf{Useless}, \textbf{Too complex} and \textbf{Resistance to change} (see Table~\ref{table:tableFearsA}).\\
        \begin{minipage}{0.40\linewidth}
            \centering
            \includegraphics[width=1.\linewidth]{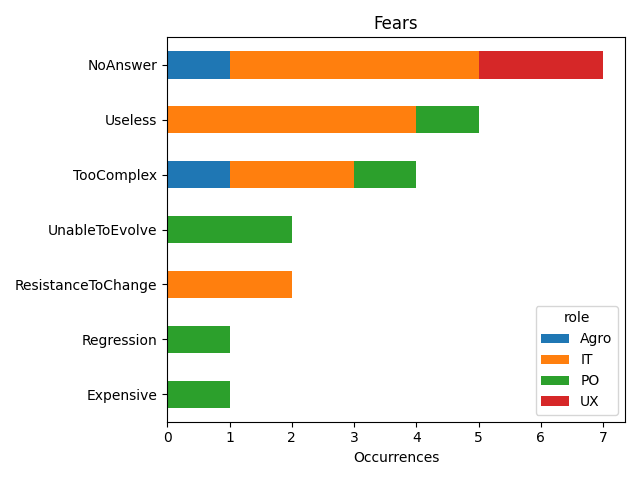}
            \captionof{figure}{The most frequently mentioned ideas about fears.}
            \label{fig:radarChartfears}
        \end{minipage}\hfill%
        \begin{minipage}{0.55\linewidth}
        \input{tableFearsA}
        \end{minipage}
        \bigskip
    
    Finally, the actors, whatever their role, asked for training 
    and demonstration. They are willing to better assess the utility of an SPL on their process to understand and get into the new approach.

%% file: tableOverviewResult.tex
\begin{table}[htb]
\centering
\caption{Results overview: number of identified current good practices, benefits, risks and fears per role and aggregated.}
\begin{tabular}{|l|r|r|r|r|r|}
\hline
                       & \multicolumn{1}{l|}{IT} & \multicolumn{1}{l|}{PO} & \multicolumn{1}{l|}{UX} & \multicolumn{1}{l|}{Agro} & \multicolumn{1}{l|}{Total} \\ \hline
Current Good Practices & 20                      & 6                       & 5                       & 3                         & 34                         \\ \hline
Benefits               & 21                      & 6                       & 5                       & 1                         & 33                         \\ \hline
Risks                  & 13                      & 6                       & 3                       & 2                         & 24                         \\ \hline
Fears                  & 8                       & 6                       & 0                       & 1                         & 15                         \\ \hline
\end{tabular}
\label{table:tableOverview}
\end{table}

%% file: tableCGPA.tex
\centering
\fontsize{6.2}{8.2}\selectfont
\captionof{table}{The most frequently mentioned ideas about current good practices per actor and aggregated.}
\begin{tabular}{|l|l|l|l|l|l|}
\hline
Role           & IT & PO & UX & AGRO & TOTAL \\ \hline
Factorization  & 5  & 1  & 2  & 1    & 9     \\ \hline
Reuse          & 5  & 1  & 2  & 0    & 8     \\ \hline
\makecell[l]{Software\\engineering\\spirit} & 4  & 1  & 1  & 1    & 7     \\ \hline
Fast           & 2  & 2  & 0  & 0    & 4     \\ \hline
Innovation     & 1  & 1  & 0  & 1    & 3     \\ \hline
Team work       & 1  & 0  & 0  & 0    & 1     \\ \hline
Testing policy  & 1  & 0  & 0  & 0    & 1     \\ \hline
\end{tabular}
\label{table:tableCGPA}

%% file: tableBenefitsA.tex
\centering
\fontsize{6.2}{8.2}\selectfont

\captionof{table}{The most frequently mentioned ideas about future benefits per actor and aggregated.}
\begin{tabular}{|l|l|l|l|l|l|}
\hline
Role              & IT & PO & UX & AGRO & TOTAL \\ \hline
Customization     & 7  & 0  & 2  & 1    & 10    \\ \hline
Fast              & 4  & 3  & 2  & 0    & 9     \\ \hline
Automation        & 3  & 2  & 0  & 0    & 5     \\ \hline
Industrialization & 3  & 0  & 1  & 0    & 4     \\ \hline
\makecell[l]{Better code\\quality} & 2  & 0  & 0  & 0    & 2     \\ \hline
\makecell[l]{Selling\\advantages} & 2  & 1  & 0  & 0    & 3     \\ \hline
No answer          & 0  & 0  & 0  & 1    & 1     \\ \hline
\end{tabular}
\label{table:tableBenefitsA}

%% file: tableRisksA.tex
\centering

\fontsize{6.2}{8.2}\selectfont
\captionof{table}{The most frequently mentioned ideas about risks per actor and aggregated.}
\begin{tabular}{|l|l|l|l|l|l|}
\hline
Role                 & IT & PO & UX & AGRO & TOTAL \\ \hline
Complex              & 8  & 2  & 0  & 0    & 9     \\ \hline
Hard to maintain      & 4  & 0  & 1  & 0    & 5     \\ \hline
Feature loss          & 1  & 0  & 0  & 2    & 3     \\ \hline
Less creativity       & 0  & 0  & 2  & 0    & 2     \\ \hline
Feature compatibility & 0  & 0  & 0  & 1    & 1     \\ \hline
Side effect           & 0  & 2  & 0  & 0    & 2     \\ \hline
Time consuming        & 0  & 2  & 0  & 0    & 2     \\ \hline
\end{tabular}
\label{table:tableRisksA}

%% file: tableFearsA.tex
\centering
\fontsize{6.2}{8.2}\selectfont
\captionof{table}{The most frequently mentioned ideas about fears per actor and aggregated.}
\begin{tabular}{|l|l|l|l|l|l|}
\hline
Role               & IT & PO & UX & AGRO & TOTAL \\ \hline
No answer           & 4  & 0  & 2  & 1    & 7     \\ \hline
Unusable           & 4  & 1  & 0  & 0    & 5     \\ \hline
Too complex        & 2  & 1  & 0  & 1    & 4     \\ \hline
Resistance to change & 2  & 0  & 0  & 0    & 2     \\ \hline
Expensive          & 0  & 1  & 0  & 0    & 1     \\ \hline
Unable to evolve     & 0  & 2  & 0  & 0    & 2     \\ \hline
Regression         & 0  & 1  & 0  & 0    & 1     \\ \hline
\end{tabular}
\label{table:tableFearsA}

%% file: 5.Discussion.tex
\section{Directions to prepare the SPL migration}
\label{sec:discussion}

    Based on the identified current good practices,  benefits, risks and fears, we provide here guidelines that we have identified to prepare and conduct the migration to an SPL. These guidelines were derived from the discussions and decisions made within the company. 
 
    In this section, we present the expected results (Sect. \ref{subsec:expect}), followed by how to exploit the current good practices (Sect. \ref{subsec:expcgp}), how to encourage future possible benefits (Sect. \ref{subsec:futureBenefits}), how to reduce risks (Sect. \ref{subsec:reduceRisks}), how to reassure fears (Sect. \ref{subsec:reassureFears}), the concrete actions (Sect. \ref{subsec:concreteActions}), and finally, a discussion (Sect. \ref{subsec:discussion}).
    
    \subsection{Expected results}
        \label{subsec:expect}
        Before starting the interviews we asked ourselves which kinds of results could be expected.
        We thought persons with knowledge or affinity with reusability to be the most positive and interested persons. They should be motivated to instigate and encourage the project implementation. In our case, the developers with the highest level of experience provided more responses related to the mindset, while the developer with less experience focused more on technical aspects. However, this variation related to the experience is not significant, in terms of the quantity or quality of the provided answers.

        Persons close to business or less technical should be the most unwilling because even admitting the utility, they could have more concerns about rapid delivery.
        As the team is relatively small-sized and used to work on new technologies, and as using an SPL was considered at the very beginning of the project, we expected low resistance to change. 
       In this context, and using the interview results, we have identified tracks (directions) to consolidate the good practices, increase benefits and limit risks and fears.
   
     \subsection{Exploiting current good practices}
        \label{subsec:expcgp}
   
         Highlighting and encouraging the existing good practices during the SPL implementation and its maintenance is a factor to reassure. 
        The process habits will evolve and a team not changing its development good practices is more likely to be motivated and to accept the novelty. The team's habits will experience significant changes, and by maintaining the current good practices, the goal is to keep the team in a familiar environment.

        \textit{Factorization, reuse and rapidity}, are three foundations of SPL~\cite{DBLP:books/daglib/0015277}.  In our context, these elements are already good practices, so we can use them to emphasize the future benefits. Migrate into an SPL is a good way to make reuse and factorization sustainable.
        
        \textit{Innovation} is already an important part of the current process. The effectiveness of SPL should encourage the adoption. 
        The current process of ITK involves an improved \textit{clone-and-own} approach, incorporating SPLE reflections, such as the \textit{``by and for reuse''} development process.
        Motivation may thus come from the fact that the time saved by giving up the time-consuming \textit{clone-and-own} approach~\cite{DBLP:conf/csmr/DubinskyRBDBC13} can be used for more innovative feature development. 
        In addition, developing in the context of an SPL may enhance team cohesion around an inspiring paradigm.

    \subsection{Encourage future possible benefits}
        \label{subsec:futureBenefits}

        In the expressed benefits we found other foundations of SPL such as \textit{Customisation}, \textit{Automation} and \textit{Industrialization} \cite{DBLP:books/daglib/0015277}. There are also other criteria which are possible to  discuss 
        and include. For example,  \textit{better code quality} can be  encouraged  with the benefit of repaying the technical debt. A code base is more homogeneous, consistent and with less bugs while avoiding the clone-and-own situations. The common bug correction also increases the code quality.
        
        The selling advantages are surprisingly not only mentioned by \textit{POs}, and are even more mentioned by \textit{ITs}.
        For example, using this kind of engineering may confirm and consolidate the research and development (R\&D) image of the company. As another example, SPL adoption promotes faster production of new software. In addition, it fosters the creation of new products with novelties for each similar business market. 
        These two examples can motivate teams in SPL adoption.
        
    \subsection{Reduce risks}
        \label{subsec:reduceRisks}
        Reducing risks is the key to limit the resistance to change. Here, we detail some tracks to reduce them.  All fears, related or not to these risks, have to be considered because they prevent from a good adoption.
        Some risks or weaknesses are recurrent in SPL adoption. 
        But they are not insurmountable if we follow and guide the migration taking them into consideration.
        
        The \textit{complexity} feeling is explained by a misconception of SPLE. But an SPL might also pose complex challenges for those who are familiar with SPLE. Complexity can exist within the SPL itself, but it should not exceed the complexity of the codebase on which the IT staff already works, and masters collectively its complexity. Within the SPL, certain assumptions about variability are made explicit through models, aiming to enhance ``understandability" and consequently address the complexity.
         We think that the request of training and demonstration could be connected to this feeling. 
        Thus we explained that in contrast, an SPL is a gain of time usable to implement new features and to perform more creative development. By detailing this, we avoid the risks of \textit{loosing creativity time} as expressed by \textit{UXs} and \textit{loosing time for innovative development}.
        \textit{Side effects} are avoidable with more systematic non-regression testing.

    \subsection{Reassure on fears}
        \label{subsec:reassureFears}
        Fears are avoidable by taking time to listen the team worries, and to discuss on weaknesses and strengths~\cite{WEIHRICH198254_SWOT}. In addition they are seen as a bias expressed by the interviewees.

        Seeing an SPL as a system which is \textit{too complex to use and evolve, making the new approach useless} can be understood. 
        Trust can be restored by giving access to the SPL as a white box, with a lot of explanations and a rich documentation, 
        and by including all possible actors in the project.
        The inclusion of the actors starts by asking them participating to the migration. 
        It is important to encourage them to make regular reporting at all the different steps.
        This will allow giving them the feeling they are listened and understood and bringing them assistance once needed.%
  
  \subsection{Concrete Actions}
    \label{subsec:concreteActions}
        The concrete actions facilitate a smoother transition and are widely employed in various software development life cycles, particularly during SPLE migration.
        Many agile practices could be used to minimize the negative effects of migration adoption. For example, a Sprint Zero~\cite{DBLP:conf/birthday/Boehm10} is a possibility to improve teamwork during the migration. This sprint contains different steps:
        The first step is building the \textit{working agreement}, where we plan the agile rituals helping team work doing the migration.
        Then having a clear and common \textit{Vision} is a prerequisite to be sure that all team members have understood the goal and the benefits of the SPL. The third step is building the \textit{persona}, i.e., the different user profiles or the different developer profiles  from the company, who are determined before starting the project to gain empathy and establish who will use it. The next step is building the \textit{Backlog} by taking into consideration all the needed work. Then each task from the backlog has to be \textit{estimated} and put in a road map.
        Planning recurrent retrospectives for the team to share their feeling is a way to improve team work. 
        
        Some software engineering practices are useful to improve the project quality, like regular refactoring sessions to increase reuse practices and remove technical debt. These practices ensure a better quality code: first plan regular testing sessions to avoid bugs, planning acceptance testing sessions with the final users is also a good way to be sure that we are on the same wavelength. Constantly monitor the code execution is a possibility to keep the derived applications as efficient and scalable as possible.
        
        Organize tool testing sessions can give an opportunity to the actors to familiarize themselves with tools and understand their capabilities~\cite{DBLP:journals/jmis/Chau96}. In the case of SPL adoption, these tools can be the clone-and-own framework Ecco~ \cite{DBLP:conf/icse/0006LLE15}, Feature IDE to build Feature Models and variants~ \cite{DBLP:journals/scp/ThumKBMSL14}, implementing SPL with Mobioos Forge~\cite{mobioos} or the framework But4reuse~ \cite{DBLP:conf/icse/MartinezZBKT17}. 
        
        These concrete actions are of course general and can be applied to any software development setting, whether it is SPL-based or not. We see strengthened advantages in an SPL-based development setting, since the improved quality is embedded at the heart of the code base. This benefits the current applications, but will also benefit for future derived applications.
        
        \subsection{Discussion}
            \label{subsec:discussion}

            The journey toward migrating to SPL extends over a significant period. It starts with an initial phase marked by prevailing practices and habits, followed by discussions to assess the viability of the migration and whether it is a prudent decision. The migration phase follows, during which we work on gathering insights and addressing resistance to change, including the migration of source code. Ultimately, the post-migration phase means the initiation of a fresh development process.

            The transition to an SPL approach entails substantial organizational changes, especially considering that Software Product Management, in itself, still presents significant challenges, as evidenced by the survey conducted by Springer and Miler \cite{DBLP:journals/ese/SpringerM22}.
             With the adoption of an SPL, teams become more cross-functional, collaborating on a shared set of core assets that are reusable across multiple projects. This shift allows for optimized resource utilization, shorter time-to-market, and a focus on innovation. The emphasis on component reusability, configuration-based customization, and unified documentation streamlines processes and enhances efficiency. While this transition requires careful planning and change management, it ultimately fosters a culture of collaboration, efficiency, and adaptability within the organization.
              
            In the context of ITK, the outcomes and guidelines were showcased and discussed upon with the development team. They expressed the desire to be well-informed and involved, and in response, we engaged in continuous collaboration and progression alongside them throughout the entirety of the migration process. We can put the findings and the reflections made at ITK in perspective with several reports of SPL adoption, strategies, and experiences of the literature.
            
           For instance, Böckle et al.~\cite{DBLP:conf/splc/BockleMKKLLNSW02, Bockle/10.1007/3-540-28901-1_2} proposed a strategy that involves four phases: Stakeholder Identification, Motivation for Product Line Adoption, Development of an Adoption Plan, and Launching the SPL. The \textit{Framework for Software Product Line Practice} introduces practices for successfully implementing a technology change along different perspectives: vision, skills, incentives, resources, and action plans. We have already identified the stakeholders, particularly those who play an active role such as Developers or Product Owners. At the project beginning, discussions were held with stakeholders from the financial and management personnel. In this study, we are currently in the stage equivalent to \textit{Motivation for Product Line Adoption}, and we are in the process of formulating the \textit{Adoption Plan}.
 
           The ITK's action plan adheres to the \textit{IDEAL improvement plan}~\cite{McFeeley/IDEAL}. The project started with the collegiate decision to \textit{initiate} the migration. Our work starts at this time with \textit{diagnosing} the practices and the project perception. The next steps of the ITK adoption are \textit{establishing} a strategy and an action plan, \textit{acting} the plan and finally, classically in an agile approach, \textit{learning} from the successes and the failures to continuously improve the development process. Our guidelines serve as a method for directing the establishment design of the action plan.
           
           There are several differences between our barriers and those identified in \cite{DBLP:conf/splc/NorthropJ13}.
            For example the \textit{Lack of necessary talent} or \textit{Lack of sufficient management support} are absent from the detected perception due to the company mindset. In \cite{DBLP:conf/splc/NorthropJ13}, an interview strategy outlined involves being aware about various aspects such as ``Background'', ``Advantages'', ``Challenges'', ``Risks'', ``Strategies'', and ``Technological Transformations''. In our interview design we share with them some elements like ``benefits'' and ``risks'', while others, such as ``barriers'' are close to our ``fears''.

%% file: 6.ValidityStudy.tex
\section{Threats to Validity}
\label{section:validity}

    In this section we report on the possible biases that we have identified in our research method. We also discuss how we mitigated them in order to provide an objective empirical study, as recommended in~\cite{DBLP:books/daglib/0029933}.
    We consider in particular threats to construct validity (Sect. \ref{subsec:construct}), internal validity (Sect. \ref{subsec:internal}), external validity (Sect. \ref{subsec:external}), and conclusion validity (Sect. \ref{subsec:conclusionv}).
    
    \subsection{Construct Validity}
    \label{subsec:construct}
        We examine here whether the questionnaires and metrics capture the concepts they are supposed to assess.
    
        We explained in Section \ref{sec:methodo} how the questions map to four axes corresponding to the research questions. 
        SPLE was unknown by most of the interviewees, so we described the SPLE concepts and purposes during the interviews.
        This involved a potential bias introduced during the explanation.
        However, we trust our interviewees to be able to form their own opinion on the subject, and the interviewer to avoid having too much impact on this opinion.

        The questions are intentionally open-ended to encourage extensive responses and capture a wide range of concepts. They have been carefully built to avoid limiting the scope extent of potential answers.
        
        The metrics were built to get the most out of the information we obtained, and to extract as much useful data as possible. 
        The metrics count the relevant terms and their occurrence number (aggregated and per role) in the answers. So tables contain exhaustive information and (bar) graphs highlight value distribution.
        Even if an answer is unique and not shared by anyone else, it is still considered and taken into account. We conduct an exhaustive reflection on all the answers provided.
        
        Terms extracted from the interviews were aggregated and aligned with terms from the SPLE domain, like ``\textit{personalisation}'' and ``\textit{making specific some parts of the code}'', which correspond to ``\textit{customization}''. 
        This helped in making uniform the results. Besides, this alignment helped finding in the literature the arguments to reassure actors worries~\cite{ESE:inTheLarge, DBLP:conf/splc/MartinezZBKT15, DBLP:journals/csur/Krueger92}.

    \subsection{Internal Validity}
        \label{subsec:internal}

        We report here possible bias in collecting and analyzing data. 
        The interviews were organized with a small panel of 15\% of the SME staff, i.e. 50\% of the IT staff, 30\% of the POs, 10\% of the agronomists and all the UI/UX Designers.
        The staff working on the project with us changed these last months, moving to other projects within the company. We have discussed with persons that were present from the starting of the development of the code base, but also with persons who came or left along the way.
        Even these percentages seem to be small, the panel of interviewees is representative of roles and of different periods in the project.
        The results of those discussions were then interpreted and analyzed by the interviewer, but they were also discussed with the other co-authors of this work in order to limit the misunderstandings.
        
        Iterating through this process with many different persons enabled us to evolve positively the fluidity with which the interviews have been conducted. Questions have not been changed, but the way questions were approached was improved over time. For instance, this involves offering additional hints and elaborating on details when the interviewee is unsure, or enhancing the transitions between questions. Despite this, we believe that, whether on the first or the last questionnaire, the quantity and quality of information extracted during these interviews were the same.

        Fears expressed by our interviewees can be perceived as a bias in the results. They are subjective feelings and depend much on the personality and experience of the project's actors. But since we interviewed 16 persons and presented aggregated answers, the bias is slightly mitigated.

    \subsection{External Validity}
        \label{subsec:external}
        In external validity, we aim to assess what can be generalized in this work to other SPL migration contexts.
         
        Our questionnaires have some company-specific questions. These are related to the development process practiced by the company. Except this, the remaining questions are related to code reuse and SPL in general. We believe that the development process of the company, which is part of a larger SCRUM-based project management methodology, is common to many companies of the same size or smaller ones. We think thereby that the results discussed previously can be generalized to other contexts of companies using agile methodologies and organizing their development teams in the same way (IT, PO, ...).
    
        Our study primarily concentrates on an SME which collaborates with our research team. All participants are from the same company and are questioned about the same SPL project. Nonetheless, we designed the process and guidelines so that they are broad enough to be applied to other SPL migration cases.
        
        Conducting personalized interviews according to the role and responsibilities of the interviewees and then performing a unique / centralized analysis of the results is a good way to get the interviewees committed and to obtain more precise answers.
        This enabled to collect a dataset of a good quality that guides the discussions on how to prepare and adapt the SPL migration to the different needs.
        As indicated previously, all this data is available online to allow the replication of our method in other contexts:\\ \href{https://gite.lirmm.fr/tgeorges/interviewartefact}{https://gite.lirmm.fr/tgeorges/interviewartefact}.
        
    \subsection{Conclusion Validity}
        \label{subsec:conclusionv}
        In conclusion validity, we evaluate to which extent we draw reasonable conclusions and deduce relevant directions for the migration and for addressing potential resistance.
        The identified good practices, benefits, risks and fears can be put into perspective with approaches to overcome resistance to change~\cite{alpha2004}: Alpha strategy consisting in advice to increase motivation and Omega strategy consisting in tips to decrease avoidance.
        Actors identified good practices that are recommended for agile projects~\cite{beck2001agile}, and this is inline with the habits of the company and with the recommendation \textit{emphasize and commitment} of the Alpha strategy.
        The identified benefits are traditional benefits for SPL and encouraging them supports another advice in Alpha strategy, which is \textit{add incentive}. 
        Similarly, the extracted risks and fears make sense. The request of different actors to have concrete solutions to reassure about these risks and fears such as training, examples and demonstrations that will serve as proofs of concept, corresponds to 
        \textit{giving guarantees and raise self-esteem} from Omega strategy. This also will \textit{improve source credibility} from Alpha strategy.

%% file: 7.RelatedWork.tex
\section{Related work}
\label{section:relatedwork}

    The stage we studied in this work is the preparation of the SPL implementation, after the company decided the adoption of this type of software engineering strategy. In this section, we successively position our  work in the fields of SPL adoption, interview design, agility and empirical software engineering.

    \paragraph{SPL adoption} This stage lies between SPL adoption convenience study, such as the one discussed in~\cite{DBLP:conf/rcis/RinconMS18, DBLP:conf/splc/RinconMS19} and  maturity or post-migration evaluation study, as reported in~\cite{ESE:inTheLarge}.
    Lindohf et al~\cite{ESE:inTheLarge} share similar research questions and describe the different perspectives and frameworks to assess SPL maturity. They also distinguish questions according to the interviewee's expertise in the company. They discussed the need for a systematic approach with a wide range of factors to assess the impact on the organization. They also assessed the interest of the Family Evaluation Framework~\textit{(FEF)}~\cite{DBLP:books/daglib/0018329}. Rinc\'on et al.~\cite{DBLP:conf/rcis/RinconMS18, DBLP:conf/splc/RinconMS19} proposed the APPLIES framework to decide the SPL adoption readiness. They also provide an assessment of the perception of this framework. They ask questions to decide the feasibility for a migration, while our research questions correspond to the stage where the migration has already been decided. One of the APPLIES research question operates as an early phase to the migration process, focusing on the evaluation of an organization's motivation and readiness to adopt a PLE approach. In the initial stages of PLE implementation, various factors shape individuals' attitudes towards the approach. Rinc\'on et al research divides the concept of ``convenience'' into motivation and preparedness, aligning with activities in the innovation decision process's initiation stage (agenda-setting and matching). 
    Hetrick et al~\cite{DBLP:conf/oopsla/HetrickKM06} propose solutions to avoid the innovative project resistance barrier, just like we are looking for. They detail how to start a chain reaction leading to the success of the migration on a large project. These reports have been inspiration sources for the questions we designed. In addition, they are complementary to our work, since we can use them to provide strategies to improve the effectiveness of adopting SPLE.
    In the literature, there are reports on the adoption and the implementation of SPLE techniques. 
    The authors in~\cite{DBLP:conf/splc/BockleMKKLLNSW02} describe a method used to make the transition from a standard development process into an SPL one. A study of the strengths, benefits and problems occurring while adopting SPLE is presented in~\cite{DBLP:books/daglib/0018329}. The motivation of adopting SPL in an agile context is developed in~\cite{DBLP:conf/splc/FogdalSK0Z16}. Cases of SPL adoption and the used methods are detailed in~\cite{DBLP:conf/models/BergerNRACW14}. In addition, a discussion of the key factors (e.g. processes, policies and resources) was developed to show how they impact SPL establishment. 
    A survey assessing variability modeling in the industrial practice using SPLE~\cite{DBLP:conf/vamos/BergerRNABCW13} studied how to fill the lack of empirical data on the validation of existing techniques. Their survey was made to identify the practices, opportunities and challenges in modeling variability and got 67 answers from companies. From the answers, the authors discuss in particular the importance of variability modeling in industry to improve quality and efficiency.

    \paragraph{Interview design}
    To create our questions we used a standard vocabulary to which we aligned the terms of the interviewees. 
    To establish this vocabulary, we used terms from~\cite{DBLP:conf/splc/MartinezZBKT15, DBLP:journals/csur/Krueger92, DBLP:books/daglib/0018329}. Martinez et al~\cite{DBLP:conf/splc/MartinezZBKT15} provide a practical and effective approach for organizations interested by adopting SPLE methods in their software development practices. They focus on evolving a set of software by introducing product line features and practices. They called this approach a \textit{bottom-up} approach in opposition with the \textit{top-down} approach used to build an SPL from scratch.
    We also enriched our study with the TOWS analysis~\cite{WEIHRICH198254_SWOT} inherited from the SWOT analysis~\cite{DBLP:journals/dss/HoubenLV99} to help decision making and project planning. The SWOT matrix consists of listing strengths, weaknesses, opportunities, and threats in four dedicated quadrants. The TOWS matrix consists of identifying relationships between these factors. These two analyses are used to develop more effective and informed strategies by taking into consideration the project strengths and weaknesses.
    We also designed our questionnaires so as to include the actors of the project, as recommended by: (1) Pikkarainen et al~\cite{DBLP:journals/ese/PikkarainenHSAS08} who explain the impacts of agile methods such as SCRUM or eXtreme programming on the communication, especially in a large project; and (2) McHugh et al~\cite{DBLP:journals/software/McHughCL12} who studied trust in an agile team. The teams in agility have to work in autonomy and face new difficulties. Both references~\cite{DBLP:journals/ese/PikkarainenHSAS08,DBLP:journals/software/McHughCL12} argue that the agile team members are central for the project success. 
    As it has been demonstrated in~\cite{DBLP:conf/aswec/MoeDD08} that identifying risks and insisting on benefits is good to encourage teams, we introduced the research questions and parts of the questionnaire dedicated to these topics. They discuss the factors contributing to the effectiveness of self-organizing teams such as the presence of a clear shared vision or a management that focuses on empowering team rather than micromanaging them.
    The most significant benefits and limitations of agility in software development is detailed in~\cite{DBLP:journals/sqj/SolinskiP16}, as for instance the importance of confidence or the employee satisfaction. They provide a framework to prioritize benefits and limitations of agile software development in practice. The framework consists of identifying agile practices, assessing their usage, identifying benefits and limitations, and prioritizing them based on their importance.
    
    \paragraph{Agility}
    A point of view on the management challenges for agile teams in an industrial case developed in~\cite{DBLP:conf/issta/JonesNL16} gave us some tips for the discussion part (e.g. job crafting to improve implication in the project). The authors studied challenges faced by small and medium-sized enterprises (SMEs) in the UK when adopting DevOps practices. We used their guidelines (e.g. effective management and organisation) for an efficient adoption of new practices.
    Our work is based on an agile driven development~\cite{beck2001agile} and especially agility in small and medium-sized companies, or SMEs~\cite{DBLP:journals/jss/SilvaNOAM14}. Silva et al~\cite{DBLP:journals/jss/SilvaNOAM14} focused their research on Product Line Scoping to define boundaries, commonalities and variability in the case of SMEs. They aim to identify challenges for SMEs as stakeholder conflicts or resource constraints. We aim to identify the strategies to overcome these challenges as formalized processes and investments in the development team.
    
    \paragraph{Empirical Software Engineering}
    Authors of the foundations of Empirical Software Engineering~\cite{10.5555/1088867} discuss the legacy of Victor R. Basili, a pioneer in Empirical Software Engineering. Their work highlights the importance of empirical studies as ours and especially the use of data and evidence to guide SE practices.
    
    As discussed above, there are a lot of works in the literature that addressed the challenges of adopting SPLE in an industrial context, and its impact on current agile development practices. These works proposed many interesting procedures to follow and strategies to adopt. They gave us good insights on how to proceed in the migration process. But, none of them did undertake an empirical assessment study similar to ours, in particular as our takes place after adoption and before implementation.

%% file: 8.Conclusion.tex
\section{Conclusion}
\label{sec:conclusion}

In this paper, we have presented the first stage of a large process that we are conducting to accompany a company in the migration of its code base towards an SPL. We have made interviews that we have designed by taking into account the specifics of the company, and in particular, the various types of roles of the project actors. Our goal being to identify relevant directions to prepare the migration, we focused on research questions that will help us to know the current good practices on which we could build the migration, the benefits and risks that actors perceive, and the fears they express. Proposing adequate answers to the risks and fears, building on perceived benefits and promoting good practices will be very valuable for a smooth transition.
We identified that the company has good habits that will help in the process and has a culture of innovation that will be favorable for the project.
Actors in their majority see benefits in the migration. Finally, the main risks are SPL complexity and its maintenance.

Currently, we are developing a micro-process for the analysis of simulators (black-box components developed by agronomists) input and output data schemata to capture variability.
We are also working on a method for the analysis of issues related to user stories in the Gitlab project of the platform in order to: 1) identify features, and their associated merge requests, commit messages and Git diffs, to 2) locate these features. We leverage the use of Formal Concept Analysis~\cite{DBLP:books/daglib/0095956} and Relational Concept Analysis~\cite{DBLP:journals/amai/HaceneHNV13} to build the SPL's feature model.
As a future work, we will continue studying the code base and the development process of the company to propose a migration plan: analysis of the code base, choice of an SPL type (annotative, compositional or hybrid), implementation and application of the extraction technique, design of a maintenance strategy in accordance with (seamless to) the current practices of the company, among other tasks.